\documentclass[twocolumn,showpacs,preprintnumbers,amsmath,amssymb,superscriptaddress]{revtex4}

\usepackage{graphicx}
\usepackage{dcolumn}
\usepackage{bm}
\usepackage{SIunits}
\usepackage{epstopdf}
\usepackage{array,color}
\usepackage{xspace}

\newcommand{\Rb}{Rb$_2$Cr$_3$As$_3$\xspace}
\newcommand{\K}{K$_2$Cr$_3$As$_3$\xspace}
\newcommand{\Cs}{Cs$_2$Cr$_3$As$_3$\xspace}

\newcommand{\RNum}[1]{\uppercase\expandafter{\romannumeral #1\relax}}

\begin{document}

\title{$^{133}$Cs and $^{75}$As NMR investigation of the normal metallic state of the quasi-one-dimensional  Cs$_2$Cr$_3$As$_3$}

\author{Haizhao Zhi}
\author{Drake Lee}
\affiliation{Department of Physics and Astronomy, McMaster University, Hamilton, Ontario L8S4M1, Canada}
\author{Takashi Imai}
\affiliation{Department of Physics and Astronomy, McMaster University, Hamilton, Ontario L8S4M1, Canada}
\affiliation{Canadian Institute for Advanced Research, Toronto, Canada M5G 1Z8}
\author{Zhangtu Tang}
\author{Yi Liu}
\affiliation{Department of Physics, Zhejiang University, Hangzhou 310027, China}
\author{Guanghan Cao}
\affiliation{Department of Physics, Zhejiang University, Hangzhou 310027, China}

\date{\today}

\begin{abstract}
We report $^{133}$Cs NMR and $^{75}$As Nuclear Quadrupole Resonance (NQR) measurements on the normal metallic state above $T_c$ of a quasi-one-dimensional superconductor  \Cs ($T_c < 1.6$~K).  From the $^{133}$Cs NMR Knight shift $^{133}K$ measured at the Cs1 site, we show that the uniform spin susceptibility $\chi_{spin}$ increases from 295~K to $\sim$ 60~K, followed by a mild suppression; $\chi_{spin}$ then levels off below $\sim$10~K.  In contrast, a vanishingly small magnitude of $^{133}K$  indicates that Cs2 sites contribute very little to electrical conduction and the exchange interactions between 3d electrons at Cr sites.  Low frequency Cr spin dynamics, reflected on $^{75}$As $1/T_1T$ (the nuclear spin-lattice relaxation rate $1/T_1$ divided by temperature $T$), shows an analogous trend as $\chi_{spin}$.  Comparison with the results of $1/T_1T$ near $T_c$ with \K($T_c=6.1$~K) and \Rb($T_c=4.8$~K) establishes a systematic trend that substitution of K$^{+}$ ions with larger alkali ions progressively suppresses Cr spin fluctuations together with $T_c$.  

\end{abstract}


\maketitle

\section{Introduction}
Since the discovery of high $T_c$ superconductivity in iron pnictides in 2008 \cite{iron}, the interplay between spin fluctuations, reduced dimensionality, and the unconventional nature of superconductivity has been attracting renewed interest.  Next to iron in the atomic periodic table is chromium; interestingly, antiferromagnetic chromium-arsenide CrAs also undergoes a superconducting transition at $T_{c} \sim 2$~K under a modest pressure of 8 kbar \cite{wu2014superconductivity,kotegawa}. More recently, Bao et al. discovered that a quasi-one-dimensional \K is also superconducting in ambient pressure with $T_c=6.1$~K \cite{K-poly}. The crystal structure of \K consists of [Cr$_{3}$As$_{3}$]$_{\infty}$ chains arranged in a triangular-lattice, separated by columns of K$^{+}$ ions, as shown in Fig.\ 1.  

Early measurements of bulk physical properties such as electrical resistivity $\rho$ \cite{Kong,Kim,K-poly}, electronic specific heat coefficient $\gamma$ \cite{Kong,Kim,K-poly}, bulk magnetic susceptibility $\chi_{bulk}$ \cite{Kong,Kim,K-poly} and other measurements \cite{Raman,usr,pangGM} suggested that \K may exhibit strongly correlated phenomena owing to Cr 3d electrons \cite{Zhang,Hu}.  The nominal valence of Cr is +2.33, assuming As$^{3-}$ and K$^{+}$ with no vacancies.  This means 3.67 electrons are present in Cr 3d orbitals that form Fermi surface sheets.  First principle calculations showed that one three-dimensional Fermi surface and two quasi-one-dimensional Fermi sheets cross the Fermi energy \cite{jiang,alemany,Subedi}.  $^{75}$As Nuclear Quadrupole Resonance (NQR) measurements of the nuclear spin-lattice relaxation rate $1/T_1$ found clear evidence that Cr spin fluctuations are enhanced toward $T_c$ and may be contributing to the superconducting mechanism \cite{K-NMR}.  

\begin{figure}[!b]
\centering
\includegraphics[width=3in]{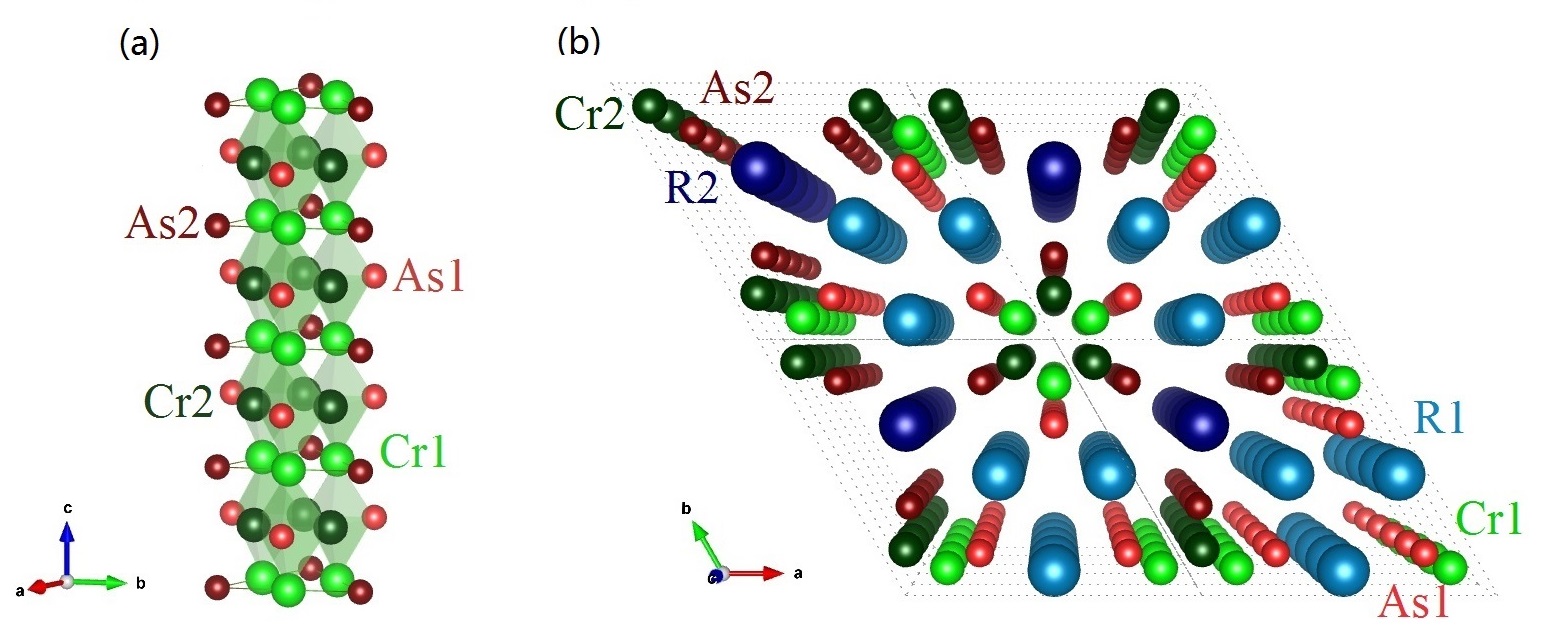}
\caption{(Color online) The crystal structure of quasi-1D superconductor system R$_{2}$Cr$_{3}$As$_{3}$ (R = K, Rb or Cs).  (a) A side view of the fundamental building block, [Cr$_{3}$As$_{3}$]$_{\infty}$ chain.  (b) The c-axis view of the crystal structure.  A complete [Cr$_{3}$As$_{3}$]$_{\infty}$ chains is in the middle.}\label{cs}
\end{figure}

Turning our attention to the superconducting properties of \K below $T_c$, the specific heat \cite{Kong,Kim,K-poly}, $^{75}$As NQR $1/T_1$ \cite{K-NMR},  the superconducting penetration depth \cite{pangGM, pang2}, and the effect of impurity scattering on $T_c$ \cite{liu2016effect} indicated a strong possibility of the unconventional nature of superconductivity.  For example, the Hebel-Slichter coherence peak of $1/T_1$, expected for the isotropic s-wave pairing of Cooper pairs, is absent just below $T_c$ \cite{K-NMR}.  In addition, the upper critical field $H_{c2} $ \cite{K-poly,Kim,Kong,Bala,zuo} has a peculiar anisotropy; $H_{c2}^{\parallel}$ along the c axis seems to show a Pauli-limiting behavior, whereas $H_{c2}^{\perp}$ along the perpendicular direction to the c-axis (i.e. normal to the chain) is much larger in the low temperature range $T \ll T_c$ \cite{Bala,zuo,Kong,Kim}. On the other hand, the in-plane $H_{c2}(\phi)$ modulation data seems to support $f$-wave coupling originating from the three dimensional $\gamma$ band \cite{zuo}.  We note that  theoretical model calculations suggest a variety of scenarios, ranging from spin-triplet superconductivity to large electron-phonon coupling BCS in \K\cite {Hu,Zhang,Subedi,zhang2015revisitation,Twisted}. 

Another interesting aspect of the \K superconductor is that one can replace the K$^{+}$ ions with other alkali ions Rb$^{+}$ and Cs$^{+}$ with larger ionic radii, and tune the superconductivity.  The a-axis lattice constant of \Rb is $\sim 3$~\% larger than \K, whereas the a-axis lattice constant of \Cs is $\sim 6$~\% larger than that of \K; the average Cr-Cr bond length within the chain, however, is almost identical for \K, \Rb and \Cs \cite{Kim,Cs-poly}.  Tang et al. showed that $T_c$ is systematically suppressed by separating the [Cr$_{3}$As$_{3}$]$_{\infty}$ chains further with larger alkali ions: $T_c = 4.8$~K for \Rb \cite{Rb-poly}, and $T_c=2.2$~K for \Cs \cite{Cs-poly}. (We should note that the superconducting shielding fraction of polycrystalline \Cs was found to be as little as 6 \%, compared to 90 \% of single crystal \K and 30 \% of single crystal \Rb \cite{Kim}.  This may be because the superconducting transition of \Cs is far more sensitive to impurity and/or disorder \cite{liu2016effect}.)  Such a systematic variation of $T_c$ naturally raises an important question: what is the key controlling factor of $T_c$?  Comparison of various physical properties reported thus far seem to indicate that many of the characteristic properties of \K are shared with \Rb and \Cs \cite{Kong,Kim,K-poly,Rb-poly,Cs-poly, Rb-NMR}.   

In this paper, we report microscopic $^{133}$Cs NMR and $^{75}$As NQR measurements of \Cs. We identify two different sets of  $^{133}$Cs NMR signals arising from Cs1 and Cs2 sites, and demonstrate very different characteristic properties between the two. The uniform spin susceptibility $\chi_{spin}$ as measured by the NMR Knight shift at the Cs1 sites show a non-monotonic temperature dependence, and is suppressed at lower temperatures. We also measured the Cr low frequency spin fluctuations through $1/T_1T$ (the nuclear spin-lattice relaxation rate $1/T_1$ divided by temperature T) at the $^{75}$As sites using Nuclear Quadrupole Resonance (NQR) in zero external field. We found that, unlike \K\cite{K-NMR} and \Rb \cite{Rb-NMR}, low frequency Cr spin fluctuation are {\it suppressed} towards $T_c$.   

\section{Experimental}

We synthesized a polycrystalline \Cs sample following the recipe described in detail by Tang et al.\cite{Cs-poly}. We sealed the powder sample in a quartz tube in $\sim 0.5$~bar of Helium gas to avoid decomposition of the sample in air and ensure good thermal contact.  The outer diameter of the quartz tube is $\sim6$~mm, and the lateral length is approximately $\sim2$~cm.  We conducted all the $^{133}$Cs NMR and $^{75}$As NQR measurements using a home-built pulsed NMR spectrometer.  

In-situ ac-susceptibility measurements conducted with a NMR coil at the radio-frequency range ($\sim$42~MHz) did not reveal any signature of superconductivity in our NMR sample down to 1.6~K.  This means that $T_c$ of the present sample is lower than 1.6~K, and/or the volume fraction is extremely low.  We recall that the volume fraction of superconductivity in the original report was as little as $\sim$6 \% below $T_c=2.2$~K  \cite{Cs-poly}.  Moreover, the superconducting properties in R$_{2}$Cr$_{3}$As$_{3}$ are known to deteriorate as the sample ages.  $T_c$ and the volume fraction of our sample may have deteriorated by the time we carried out NMR measurements. 

\section{$^{133}{Cs}$ NMR results}

We conducted $^{133}$Cs NMR measurements in an external magnetic field of $B_{o} = 9$~T around the bare resonance frequency $f_{o}=(\gamma_n/2\pi)B_0=50.26$~MHz, where the nuclear gyromagnetic ratio of $^{133}$Cs is $^{133}\gamma_{n}/2\pi = 5.58469$~MHz/T, and the $^{133}$Cs nuclear spin is $7/2$.  We present a representative frequency-swept $^{133}$Cs NMR lineshape observed at 77~K in Fig.\ 2; we found two prominent peaks around 50.5~MHz and 50.3~MHz.  The interval between successive spin-echo sequences for this measurement was 3.9~s; the smaller peak around 50.3~MHz is suppressed by about 20\% because of the extremely slow spin-lattice relaxation time $T_1$.  We assign the larger peak to the Cs1 site, because the abundance of the Cs1 site is three times larger than that of the Cs2 site in \Cs.  The powder averaged  lineshape of the Cs1 site exhibits noticeable shoulders arising from a small nuclear quadrupole interaction frequency $^{133}\nu_{Q}\sim 60$~kHz.  The $^{133}\nu_{Q}$ of the Cs2 sites appears even smaller, and we were unable to resolve the shoulders. 

These narrow powder NMR lineshapes allowed us to use the Fast Fourier Transform (FFT) of the envelope of the spin echo signals to accurately determine the peak resonance frequency $f$ at each temperature.  We summarize the temperature dependence of the NMR Knight shift, $^{133}K = (f-f_{o})/f_{o}$, in Fig.3.  Quite generally, the Knight shift may be represented as 
\begin{equation}
^{133}K = \sum_{i} \frac{A_{hf}^{(i)}}{N_{A}\mu_{B}} \chi_{spin}^{(i)} + K_{chem},
\end{equation}
where $i$ is the band index, $A_{hf}^{(i)}$ is the hyperfine coupling between the $^{133}$Cs nuclear spin and the relevant electrons in the i-th band, $N_{A}$ is  Avogadro's number, $\mu_{B}$ is the Bohr magneton; $\chi_{spin}^{(i)}$ is the i-th contribution to the overall bulk spin susceptibility $\chi_{spin} = \sum_{i} \chi_{spin}^{(i)}$, and $K_{chem}$ is the small chemical shift associated with the orbital motion of the electrons.    

\begin{figure}

\centering
\includegraphics[width=3in]{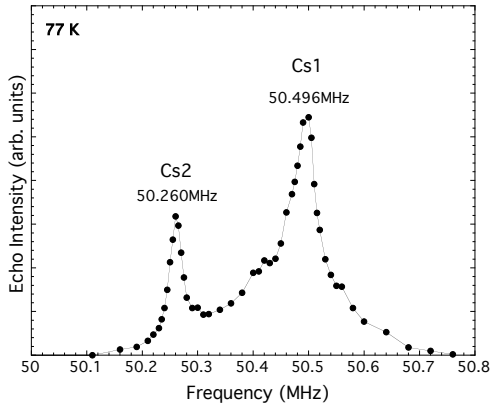}
\caption{The frequency swept $^{133}$Cs NMR lineshape of the Cs sites measured in 9~T at 77~K with the pulse separation time $\tau=60~\mu$s.  The spin echo sequence was repeated with an interval of 3.9~s.}\label{cs}
\end{figure}

The small, temperature independent $^{133}K$ observed at the Cs2 site suggests that: (a) $K_{chem}\sim$0.02 \%, and (b) the spin contribution to $^{133}K$ is negligibly small at the Cs2 sites.  The latter implies that Cs$^{+}$ ions are merely filling the space within the crystal structure, and the electrons at Cs2 sites contribute very little to the conduction bands, and to the inter-chain and intra-chain exchange interactions between Cr spins. This finding is consistent with the fact that the spin-lattice relaxation time $T_1$ is extremely long at the Cs2 peak, as mentioned above.

In contrast, the Cs1 NMR peak is strongly shifted to $\sim$50.5~MHz, and the magnitude of $^{133}K$ reaches as large as $\sim0.5$~\%.  Since the central peak arising from the nuclear spin $I_{z} = +1/2$ to $-1/2$ transition is nearly symmetrical and shows very little signature of anisotropic line broadening, the dominant contribution to the hyperfine coupling must be the isotropic Fermi's contact interaction with the 6s electron.  In other words, the strong isotropic Knight shift $^{133}K$ observed for the Cs1 sites implies that the 6s electrons at Cs1 sites develop non-negligible spin polarization in $B_{o} = 9$~T induced by the hybridization between the Cs1 6s and Cr 3d orbitals; $^{133}K$ reflects the spin polarization at Cr sites.  For comparison, it is worth noting that  $^{133}K$ is two orders of magnitude smaller in CsC$_{60}$ \cite{stenger1994nmr}.

The $^{133}K$ at Cs1 site shows a non-monotonic behavior. $^{133}K$ mildly increases when temperature goes down.
Upon further cooling, however, the Knight shift starts to decrease, before it levels off below $\sim$10~K towards $T_c$.  The observed behavior is qualitatively similar to that observed for other unconventional superconductors  Sr$_2$RuO$_4$ \cite{PhysRevLett.81.3006} and UPt$_3$ \cite{PhysRevLett.77.1374}. 


Within the canonical Fermi liquid picture, the Knight shift is proportional to the electronic density of states (DOS) at the Fermi surface; our results might imply that the DOS at the chemical potential exhibits a non-monotonic change with temperature due to extremely sharp fine structures in the DOS, and/or the effective mass is mildly enhanced below room temperature.  Our finding of the suppression of $^{133}K$ below $\sim 60$~K may also imply that antiferromagnetic correlation grows and suppresses the uniform spin susceptibility with decreasing temperature.  But this scenario seems somewhat unlikely, because low frequency spin fluctuations are also suppressed below $\sim 60$~K, as described in section V below.  

Superposition of the $I_{z} = +1/2$ to $-1/2$ central transition by other transitions makes it difficult to measure Cr spin dynamics accurately using the $^{133}$Cs NMR peaks.  We defer our discussions on spin dynamics to the $^{75}As$ NQR $1/T_1$ results in section V.  

\begin{figure}
\centering
\includegraphics[width=3in]{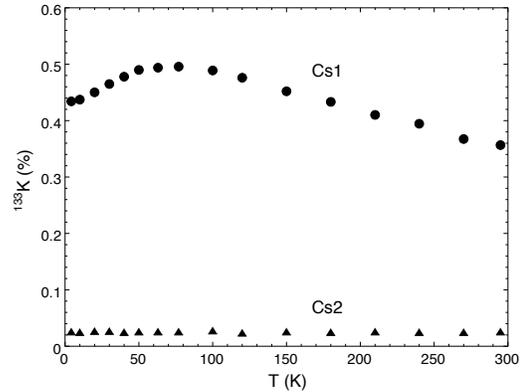}
\caption{The $^{133}$Cs NMR Knight shift $^{133}$K measured for the two different Cs sites.} \label{knight}
\end{figure}

\section{$^{75}As$ NQR lineshapes}

In Fig.4, we present $^{75}$As (nuclear spin $I=3/2$) NQR lineshapes observed between the nominal $I_{z}=\pm 3/2$ and $I_{z}=\pm 1/2$ states in zero external magnetic field.  We observed 2 peaks, called A and B hereafter.  The integrated intensity of the two peaks is comparable, as expected from the equal abundance of the As1 and As2 sites.  Both A and B peaks exhibit a broadening at lower temperatures presumably due to the strain in the lattice, which is common in many materials.  The B peak becomes noticeably broader than the A peak at 4.2~K, indicating that the former is more susceptible to disorder caused by Cs vacancies.  Since the Cs vacancies are located primarily at the Cs2 sites \cite{PhysRevB.91.180404}, the proximity between the As2 sites and Cs2 sites suggests that the B peak probably arises from the As2 sites.     

We summarize the NQR peak frequency $^{75}\nu_{Q}$ of the A and B peaks in Fig.\ 5.  The $^{75}\nu_{Q}$ of the A peak in \Cs is somewhat smaller than that of the A peak in \K, while the $^{75}\nu_{Q}$ of the B peak is comparable to that of the corresponding peak, B$_1$ \cite{K-NMR}.  Unlike the case of \K, however, we did not observe an additional side peak ``B$_2$" \cite{K-NMR} on the higher frequency side of the B peak, presumably because this \Cs sample is less disordered than the \K sample studied in our earlier work \cite{K-NMR}.    

\begin{figure}
\centering
\includegraphics[width=2.8in]{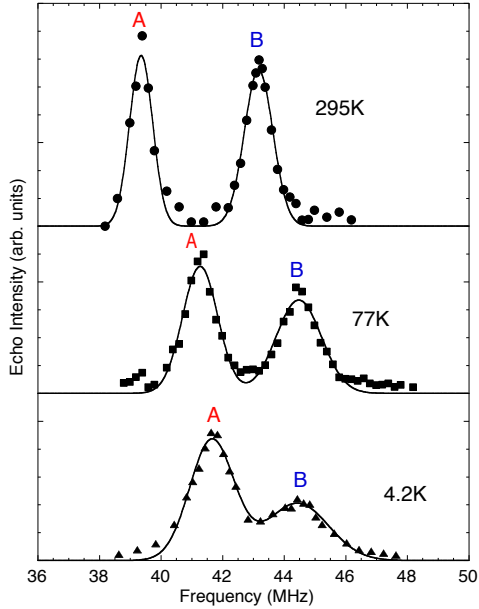}
\caption{(Color online) Representative $^{75}$As NQR line shapes. The resonance frequency increases with decreasing temperature.  The solid curves are a fit with two Gaussian functions.} \label{lineshape}
\end{figure}

\begin{figure}
\centering
\includegraphics[width=3in]{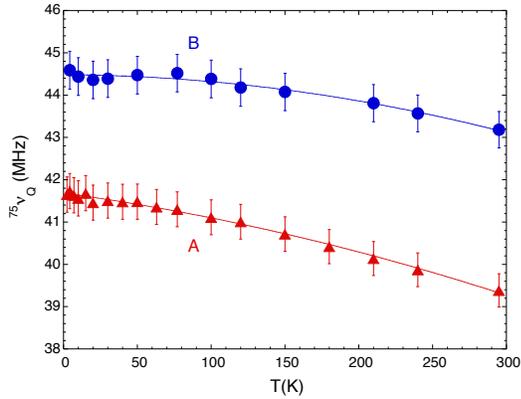}
\caption{(Color online) $^{75}$As NQR resonance peak frequency $^{75}\nu_{Q}$ as a function of temperature. The $^{75}\nu_{Q}$ gradually decrease with temperature, a typical behavior associate with thermal expansion of the lattice.  Solid curves are a guide for the eyes.} \label{lineshape}
\end{figure}

\section{Spin dynamics}

We measured the $^{75}$As nuclear spin-lattice relaxation rate $1/T_1$ at $^{75}\nu_{Q}$ by applying an inversion pulse prior to the spin echo sequence.  We fitted the observed nuclear spin recovery to the exponential form expected for the $I_{z}=\pm 3/2$ to $I_{z}=\pm 1/2$ transition,
\begin{equation}
M(t)=M({\infty})-A \times exp(-3t/{T_1}),
\end{equation}
where $M({\infty})$, $A$, and $1/T_1$ are the free parameters.  We present representative recovery curves in Fig.\ 6.  The fits were always satisfactory at 4.2~K or above, but we found a significant distribution of $1/T_1$ at lower temperatures; we will revisit this issue in section VI.  

We summarize the temperature dependence of $1/T_1$ and $1/T_{1}T$ ($1/T_1$ divided by temperature $T$) in Fig.\ 7 and Fig.\ 8, respectively.  

\begin{figure}
\centering
\includegraphics[width=3in]{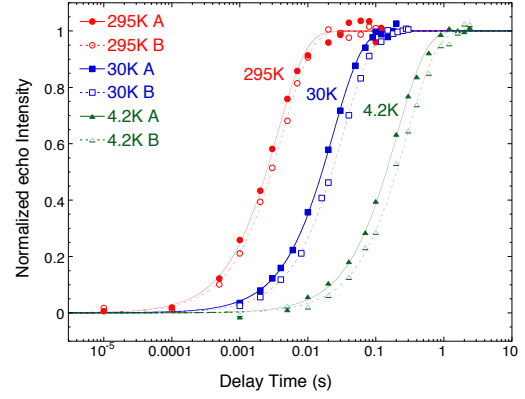}
\caption{(color online) The scaled recovery curves of both A and B peak of $^{75} As$ after an inversion pulse observed at 295~K, 30~K and 4.2~K.  Solid and dashed lines are the best fits with the Eq. (2).} \label{recovery}
\end{figure}

\begin{figure}
\centering
\includegraphics[width=2.7in]{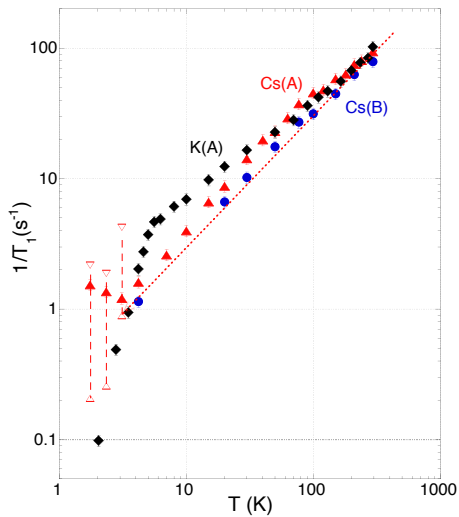}
\caption{(color online) $1/T_1$ vs temperature is plotted for \Cs in a log-log scale for both $^{75}$As A and B sites.  For comparison, we show the results for the A peak of \K. We can see clear deviation from the Korringa behavior (the dotted line, $1/T_{1} \propto T$, expected for a Fermil liquid with a broad band) for both cases, except below $\sim 10$~K for \Cs. The vertical dashed lines below 3~K mark the large distribution of $1/T_1$ in this region (see the main text for details).} \label{log}
\end{figure}

\begin{figure}
\centering
\includegraphics[width=3in]{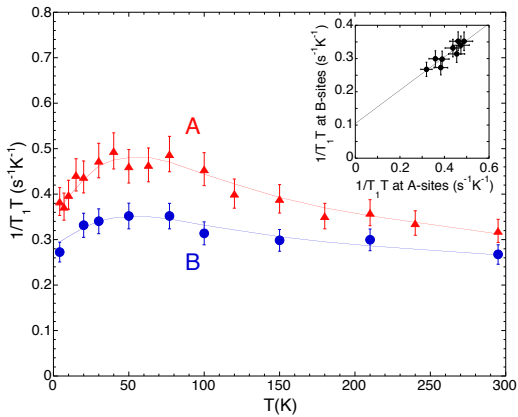}
\caption{(color online) The temperature dependence of $1/T_1T$ of A and B peak of \Cs.  We can see that A and B peak follows an analogous trend.  Solid curves are a guide for eyes.  Inset: comparison of $1/T_1T$ measured at A and B peaks, plotted with temperature as the implicit parameter.  The solid line is the best linear fit, with an offset of $\sim 0.1$~s$^{-1}$K$^{-1}$.} \label{1T1T}
\end{figure}

Theoretically, $1/T_{1}T$ may be expressed as
\begin{equation}
\frac{1}{T_1T}=\frac{2k_B}{\hbar}\sum_{i}\sum_{{\bf q}}|A_{\text{hf}}^{(i)}({\bf q})|^2\frac{\textmd{Im}\chi({\bf q},\omega)^{(i)}}{\hbar^2\omega},
\end{equation}
where Im$\chi({\bf q},\omega)^{(i)}$ is the i-th component of the imaginary part of the dynamical electron spin susceptibility (corresponding to $\chi_{spin}^{(i)}$ in Eq. (1)), $A_{\text{hf}}^{(i)}({\bf q})$ is the {\bf q}-dependent hyperfine form factor, and the wave-vector {\bf q} integral is taken over the first Brillouin zone.  Our results in Fig.\ 8 indicate that spin fluctuations mildly increases from room temperature to $\sim$60~K, then decreases toward the base temperature.  The observed non-monotonic temperature dependence is qualitatively similar to the result of the uniform spin susceptibility $\chi_{spin}$ as measured by $^{133}K$.   

Also plotted in the inset of Fig.\ 8 is $1/T_{1}T$ at B sites as a function of $1/T_{1}T$ at A sites choosing temperature as the implicit parameter; within experimental uncertainties, A sites and B sites reveal the same temperature dependence, but B sites seem to have a constant offset of $\sim 0.1$~s$^{-1}$K$^{-1}$.  This might be an indication that one (or more) of the electronic bands has a Korringa component, $1/T_{1}T \sim constant$, expected for a canonical Fermi liquid with a broad band.  The constant term appears to be superposed by an additional temperature dependent component(s) from a different band(s).  We recall that the additional $B_{2}$ sites observed for a disordered \K sample exhibited $1/T_{1}T \sim constant$ behavior in the entire temperature range \cite{K-NMR}.  Taken together, at least one of the As1 or As2 sites in the crystal structure shown in Fig.\ 1 may have a strong hybridization with a band that is highly sensitive to disorder and exhibit a Fermi-liquid like behavior.        

In Fig.\ 9, we compare the behavior of low-frequency Cr spin fluctuations above $T_c$ measured at the A sites in \K ($T_{c} = 6.1$~K) \cite{K-NMR}, \Rb ($T_{c} = 4.8$~K) \cite{Rb-NMR}, and \Cs (this work).  In the case of \K, $1/T_1T$ slowly grows toward $T_c=6.1K$, and follows a simple power-law behavior $1/T_1T \sim T^{-0.25\pm0.03}$.  The observed power-law behavior is consistent with the Tomonaga-Luttinger liquid (TLL) \cite{K-NMR} as initially suggested in \cite{K-poly} based on the $T$-linear behavior of $\rho$.  If we apply the theoretical framework of the TLL to the $1/T_{1}T$ data, the observed exponent implies that electron-electron correlation is dominated by finite {\bf q}-modes \cite{K-NMR}.  That is, if we naively accept the TLL picture, our earlier finding on $1/T_1T$ in \K implies enhancement of antiferromagnetic spin correlations.  In an independent work, it was shown that the bulk susceptibility $\chi_{bulk}$ measured for a single crystal sample of \K \cite{Kong} also grows toward $T_c$.  This means that the {\bf q} = {\bf 0} component of spin susceptibility grows in at least one band.  Analogous enhancement of both the uniform spin susceptibility as measured by $^{75}$As NMR Knight shift and spin fluctuations $1/T_1T$ were reported for \Rb, and was attributed to a Curie-Weiss enhancement of ferromagnetic spin fluctuations in \Rb \cite{Rb-NMR}.      

\begin{figure}

\centering
\includegraphics[width=3in]{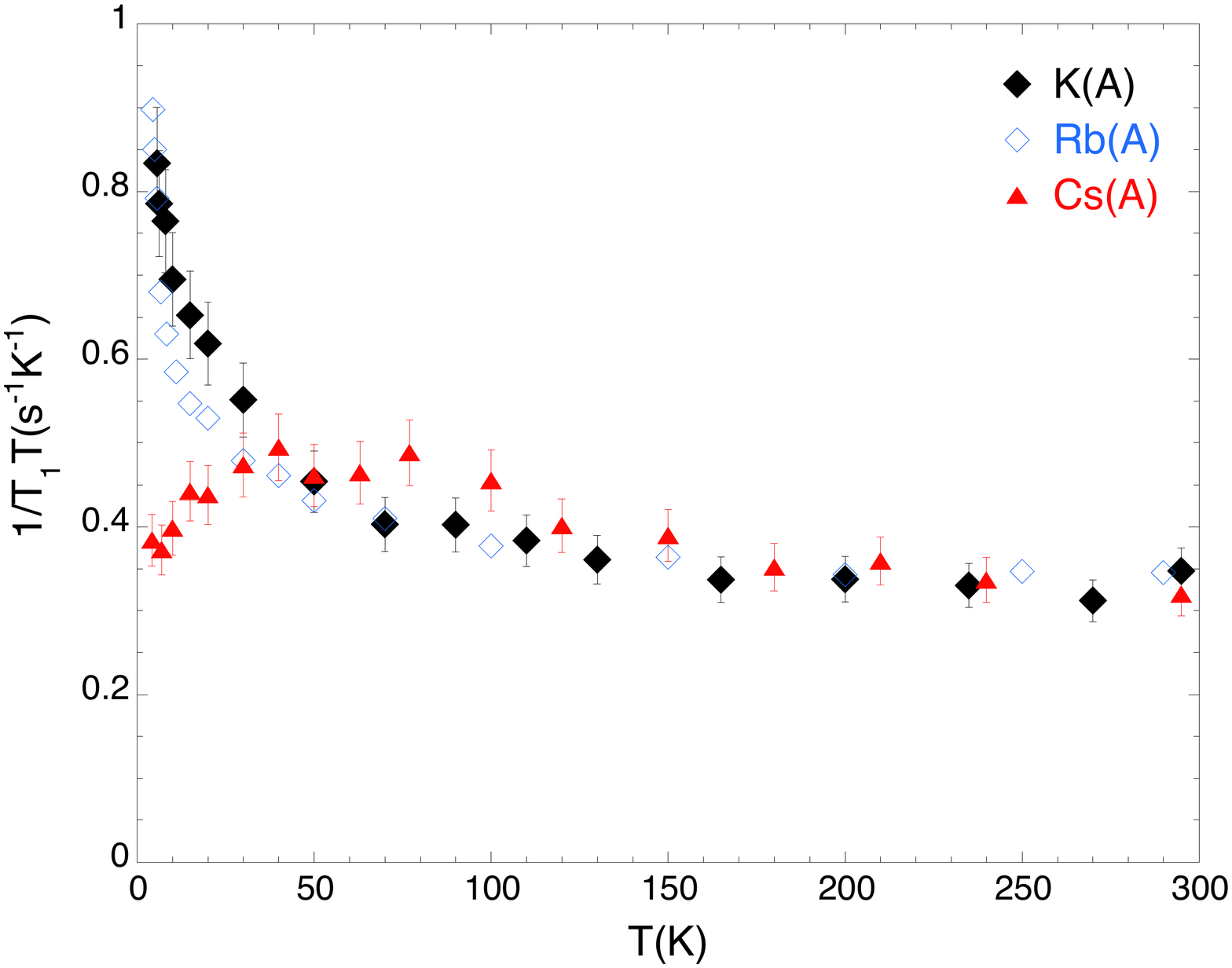}
\caption{(color online) A comparison of $1/T_1T$ observed for the $^{75}$As sites in  \K \cite{K-NMR}, \Rb \cite{Rb-NMR} and \Cs (this work) from room temperature down to their $T_c$.}
\label{K-Rb-Cs}
\end{figure}

It is useful to recall, however, that determining the nature of spin correlations in a multi-band system is not a trivial task, especially when quasi-one-dimensional bands exist.  In the case of Sr$_2$RuO$_4$ \cite{PhysRevLett.81.3006}, both the NMR Knight shift and $1/T_{1}T$ grow and saturate near $T_c$, which was initially attributed to the orbital dependent enhancement of ferromagnetic spin correlations in three different bands consisting primarily of Ru 4d$_{xy}$, 4d$_{yz}$, and 4d$_{zx}$ orbitals. Subsequently, inelastic neutron scattering measurements demonstrated that two quasi-one-dimensional bands consisting primarily of Ru 4d$_{yz}$ and 4d$_{zx}$ orbitals exhibit enhancement of spin fluctuations for finite {\bf q}-modes owing to the strong Fermi surface nesting effects \cite{PhysRevB.66.064522}.  That is, antiferromagnetic spin fluctuations are present, too, in Sr$_2$RuO$_4$.  It remains to be seen an analogous scenario may apply in the present case. 

Regardless of the exact origin(s) of the enhancement of spin fluctuations toward $T_c$ in \K and \Rb, our new data indicate that such an enhancement is absent in \Cs.  It is equally interesting to note that the strength of spin fluctuations near $T_c$ shows a systematic trend: $1/T_1T$ observed at 10~K decreases from $\sim 0.7$~s$^{-1}$K$^{-1}$ in \K to $\sim0.6$~s$^{-1}$K$^{-1}$ in \Rb, and then $\sim 0.4$~s$^{-1}$K$^{-1}$ in \Cs, while $T_c$ changes from 6.1~K, to 4.8~K, and then to $< 2.2$~K.  Such a systematic trend seems to suggest that Cr spin fluctuations are indeed linked with the superconducting mechanism, even if it is not playing a vital role.  

This systematic trend observed for spin fluctuations also seems to correlate with the variation of the electronic specific heat coefficient; $\gamma\sim73$~mJ/(K$^2$mol-f.u.) for \K \cite{Kong}, whereas $\gamma\sim 39.5 - 55.1$~mJ/(K$^2$mol-f.u.) for \Rb \cite{Kim,Rb-poly}, and $\gamma\sim 39$~mJ/(K$^2$mol-f.u.) for polycrystalline \Cs \cite{Kim,Cs-poly}.

We can test the canonical Fermi liquid picture in Fig.10, where we  plot $(1/T_1T)^{1/2}$ as a function of Cs1 Knight shift $^{133}K$ using temperature as the implicit parameter. If a canonical Fermi liquid picture applied to this case, we would observe a straight line. We observe, however, a subtle but clear deviation from a straight line in a broad range of temperature above $\sim 60$~K.

\begin{figure}
\centering
\includegraphics[width=3in]{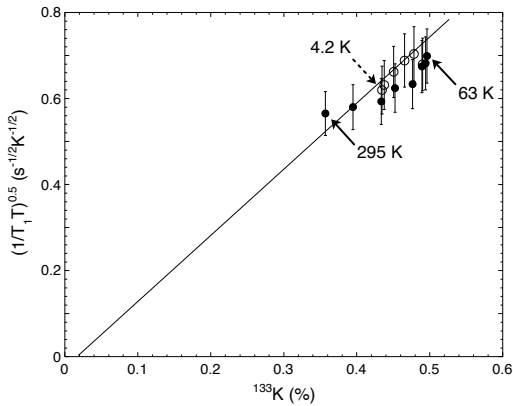}
\caption{$(1/T_1T)^{1/2}$ at $^{75}$As A sites vs the Knight shift at the Cs1 sites.  The horizontal intercept of the straight line represents the chemical shift of $^{133}$Cs Knight shift, $^{133}K_{chem} = 0.02$~\%, as estimated from the observation for the Cs2 sites.  Open (closed) symbols represent the data points below (above) 63~K.}
\label{T1TK}
\end{figure}

\section{low temperature anomaly}
Finally, we briefly discuss the results of $1/T_1$ in the low temperature region.  Above 4.2~K, the nuclear spin recovery after an inversion pulse was reproduced very well with a single exponential function in Eq.(2); the magnitude of $1/T_1$ has a negligible level of distribution.  Fitting the recovery data to a stretched exponential form with an additional phenomenological exponent $\beta$, $M(t)=M({\infty})-[M({\infty})-M(0)]exp[-(3t/{T_1})^{\beta}]$, hardly changes our results, because the value of $\beta$ hovers around $\beta= 1$ above 4.2~K, as shown in the main panel of Fig.\ 11.  

In the low temperature range below 4.2~K, we found that $1/T_1$ develops a significant distribution, and $\beta$ deviates noticeably from 1.  For example, as shown in the inset of Fig.\ 11, we found $\beta = 0.8$ at 3.1~K.  The observed distribution of  $1/T_1$ implies that our \Cs sample develops magnetic inhomogeneity at low temperatures, presumably because of the disorder caused by Cs vacancies, in analogy with the case of \K with a high level of K-deficiency \cite{PhysRevB.91.180404}.  This finding is not surprising in view of the fact that the volume fraction of superconducting shielding of \Cs is as little as 6\% at 1.9~K \cite{Cs-poly}, and the present sample does not reveal the signature of superconductivity down to 1.6~K.  The data points below 4.2~K presented in Fig.\ 7 are the results of the stretched exponential fit (filled symbols); we also show the fast and slow components of double exponential fits, connected by a dashed line, to show the range of the distributed $1/T_1$. 

\begin{figure}
\centering
\includegraphics[width=3in]{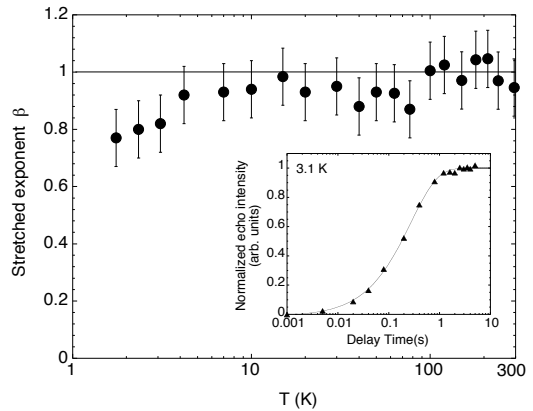}
\caption{Main panel: the temperature dependence of the phenomenological parameter $\beta$ for a fit with a stretched exponential function.  $\beta = 1$ implies that there is no distribution in the relaxation time $T_1$ and the low frequency magnetic response of the sample is homogeneous.  Inset: the recovery curve at 3.1~K, fitted with a stretched exponential function with $\beta = 0.8$.}
\label{steprecovery}
\end{figure}

\section{Summary}
In this paper, we presented the new $^{133}$Cs NMR and $^{75}$As NQR results for \Cs, and compared the results with earlier reports for \K \cite{K-NMR} and \Rb \cite{Rb-NMR}.  We demonstrated that strong enhancement of Cr spin fluctuations at low temperatures is absent in the powder sample of \Cs used for our NMR experiments.  The underlying cause of our observation is not clear.  Naively, since the ionic radius of Cs$^{+}$ ion is larger and the lattice expands along the ab-axes, one might expect that the electronic coupling between  [Cr$_{3}$As$_{3}$]$_{\infty}$ chains becomes weaker, and hence the quasi-one-dimensional nature may manifest even more strongly.  If the Tomonaga-Luttinger picture applied in the whole series of \K, \Rb, and \Cs (as initially suggested for \K by early observation of the $T$-linear resistivity $\rho$ \cite{K-poly}, and subsequently from a power-law divergent behavior of $1/T_{1}T$ \cite{K-NMR}), then enhanced Fermi surface nesting effects within the quasi-one-dimensional bands should cause additional enhancement of spin fluctuations in \Cs.  We did not observe evidence for such enhanced quasi-one-dimensional spin fluctuations.  

Instead, our new finding seems to suggest that the extended nature of the 6s orbital at the Cs1 sites might actually enhance the inter-chain coupling and weaken the correlation effects.  Perhaps a naive quasi-one-dimensional picture, and hence the Tomonaga-Luttinger picture, is not valid in these materials.  After all, a three-dimensional electronic band also exists.  

Alternately, Cs$^{+}$ ions might influence the nature of the intra-chain spin correlation effects through very subtle changes in the structure along the chain.  For example, it is conceivable that a subtle change in the balance between the nearest-neighbor and next-nearest-neighbor Cr-Cr exchange interactions along the chain might strongly affect the magnetic behavior \cite{WuXX}.  Then enhanced antiferromagnetic correlations along the chain might cause the observed suppression of $\chi_{spin}$ below $\sim 60$~K.  However, suppression of $1/T_{1}T$ below below $\sim 60$~K is contradictory to such a scenario, because $1/T_{1}T$ should grow below $\sim 60$~K.

\section{Acknowledgement}
The work at McMaster was supported by NSERC and CIFAR.   The work at Zhejiang was supported by the Natural Science Foundation of China (No. 11190023) and the National Basic Research Program (No. 2011CBA00103).\\

\bibliographystyle{unsrt}


\end{document}